\documentclass[showpacs,aps,prd,nofootinbib,floatfix,amsmath,amssymb]{revtex4}
\usepackage{graphicx}
\usepackage[utf8x]{inputenc}
\usepackage{multirow}
\usepackage{color}
\usepackage{hyperref}

\begin{document}



\title{Rare decays of the top quark  mediated by  $Z^\prime$ gauge bosons and flavor violation\\}

\author{\footnotesize J I Aranda$^1$, D Espinosa-G\'{o}mez$^1$, J Monta\~{n}o$^{1,2}$, F Ram\'{i}rez-Zavaleta$^1$,  E S Tututi$^1$ }

\affiliation{$^1$Facultad de Ciencias F\'{i}sico Matem\'{a}ticas,
Universidad Michoacana de San Nicol\'{a}s de Hidalgo, Avenida Francisco J. M\'{u}gica S/N, 58060\\
Morelia, Michoac\'{a}n, M\'{e}xico\\
tututi@umich.mx\\
$^2$$^{2}$CONACYT, Av. Insurgentes Sur 1582, Col. Cr\'edito Constructor, Alc. Benito Ju\'arez,
C.~P. 03940, Ciudad de M\'exico, M\'exico.}




\begin{abstract}
The rare top quark decays mediated by a new neutral massive gauge boson that is predicted in models with extended gauge symmetries are studied. We focus on the processes $t\to cV, uV$ induced at the one loop level, where $V =\gamma, g$, by considering different extended models. It is found that, within a broad range of mass of the new neutral gauge boson, the models predict branching ratios for the decays in study that are competitive with respect to the corresponding branching ratios in the standard model. In order to establish bound on our branching ratios, we consider the recent experimental bounds as $m_{Z^\prime}\geq$ 3.8-4.5 TeV, depending on the  model, which also impose restrictions on   our calculation. Even in this case, the resulting branching ratios  are of the same order of magnitude as that predicted by the standard model. It should be noted that  for the case of two models studied here, since no experimental bound exists to compare with, the results could be important, as they are, in the best of cases, two orders of magnitude larger than the predicted by the standard model.

\end{abstract}

\maketitle

\section{Introduction}
It is well known that the top quark decays almost exclusively into a bottom quark and  a charged gauge boson, $t\to b\,W$~\cite{qtop}, leaving little room to others modes of decay. In the literature these other modes are named rare decays, which have been of much interest and effort of analysis since they open a window to explore effects of new physics beyond the standard model (SM). Within the framework of the SM there are some modes of decay of the top quark to two or three  bodies that could occur either at the tree level or at the one-loop level, although they are very suppressed,  have sizable branching ratios \cite{diaz-eilam,jenkins,JAF}. In fact, at the one-loop level, the top quark could violates flavor symmetry through the  $t\to u_jV$ ($u_j=u,c;\,\,V = g, \gamma, Z, H$)~\cite{BM,GEilam}, $t\to cgg$~\cite{GEilam} and $t\to c\bar{c}c$ decays~\cite{ACordero}; these processes are strongly suppressed by the Glashow-Iliopoulos-Maiani mechanism. On the other hand, the $t\to c V$ decays  have been widely studied in several extensions of the  SM. For example, in Two Higgs Doublet Models (THDM)~\cite{WS,JDiaz}, in Supersymmetric Models~\cite{CS,JMYang}, in \textbf{331} models~\cite{Acid} and independent descriptions of the model~\cite{FAguila,AC}, among others.

Flavor changing neutral currents (FCNC) in the top quark sector involve transitions that are important for searching new physics effects, since these are more evident in those process that are strongly suppressed or forbidden in the SM. In the quark sector of the SM, flavor changing neutral transitions arise at the one loop level. However, in theories beyond the standard model, FCNC effects can appear at tree level in the presence of a neutral massive gauge boson, identified in a generically manner as $Z^{\prime}$ \cite{JPA,kim-langacker-levine,RJL, durkin-langacker,barger-et-al,FPV,perez-soriano}. The simplest extended model that predicts the existence of a  $Z^{\prime}$ boson  is  based on the extended electroweak gauge group $SU_L(2)\times U_Y(1)\times U^{\prime}(1)$~\cite{plangacker,leike,robinett,Plm1,arhrib1}. The search for these extra neutral gauge bosons at the Large Hadron Collider (LHC) is of  current interest. In fact, the latest results on the search for new spin-1 resonances impose updated lower limits on their masses~\cite{ATLAS2017,CMS2018}. The ATLAS Collaboration reports, with a $95\%$ confidence level in proton-proton collisions at $\sqrt{s}= 13$ TeV, the following bounds~\cite{ATLAS2017}: $m_{Z^{\prime}}> 4.5$ TeV for a sequential $Z^{\prime}$ boson, $m_{Z^{\prime}}> 4.1$ TeV for the $Z_{\chi}$ boson, $m_{Z^{\prime}}> 3.9$ TeV for the $Z_\eta$ boson and $m_{Z^{\prime}}> 3.8$ TeV for the $Z_{\psi}$ boson. The CMS Collaboration~\cite{CMS2018} communicates, for the $Z_S$ boson, a lower mass limit $m_{Z^\prime}>4.5$ TeV, and for the $Z_{\psi}$, the respective reported bound is $m_{Z^{\prime}}>3.9$ TeV.

In this work we study the effects of flavor violation in the quark sector through the $t\to u \gamma, c\gamma, ug, cg$ processes  mediated by a $Z^{\prime}$ boson. In particular, we are interested in computing the branching ratios of these rare decays and to contrast them with the related SM predictions. Accordingly, the SM predictions in question are Br$(t\to u\gamma)\sim 10^{-16}$, Br$(t\to c\gamma)\sim 10^{-14}$, Br$(t\to ug)\sim 10^{-14}$, Br$(t\to cg)\sim 10^{-12}$~\cite{JAS,geli}. On the other hand, in several extended models, the branching ratios for FCNC in top quark decays can be many orders of magnitude larger than the  predicted by the SM, as is presented as follows, where we only show results of some representative models. The THDM~\cite{geli,DAT,martinez}, gives the results: Br$(t\to c\gamma)\sim 10^{-12}-10^{-6}$ and Br$(t\to cg)\sim 10^{-8}-10^{-4}$. The Minimal Supersymmetric Standard Model (MSSM)~\cite{chong} predicts branching ratios for the $t\to c\gamma$ and $t\to cg$ decays of the order of $10^{-8}$ and $10^{-6}$, respectively. Regarding the R-parity violating supersymmetric models (RVSSM)~\cite{JM}, the Br($t\to c\gamma$) is of order of $10^{-5}$ and the Br($t\to cg$) is of order of $10^{-3}$. The Littlest Higgs Model with T-parity (LHMT) gives the following results: Br$(t\to c\gamma) \sim 10^{-7}$, Br$(t\to cg) \sim 10^{-2}$~\cite{HO}, and Br$(t\to c\gamma)<10^{-9}$, Br$(t\to cg)<10^{-8}$~\cite{xiao}. In relation to the Left-Right Supersymmetric Model (LRSSM)~\cite{MFR}, the Br($t\to c\gamma$) is of order of $10^{-6}$ and the Br($t\to cg$) is of the order of $10^{-4}$. As far as  the Extra Dimensional Model (EDM) is concern, it is reported the following results: Br$(t\to c\gamma)\sim 10^{-6}$~\cite{TIE}, Br$(t\to cg)\sim 10^{-5}$~\cite{TIE}, and Br$(t\to c\gamma) \sim 10^{-10}$~\cite{GON}. There are also estimates on the branching ratios for the decays $t\to u\gamma$ and $t\to ug$ being Br$(t \to u\gamma)<4\times 10^{-4}$ and Br$(t\to ug)<2\times 10^{-3}$, respectively, that were made in the context of model independent approach~\cite{jarandaA}.

The ATLAS and CMS collaborations have established bounds on the strength of the $tu\gamma$, $tug$, $tc\gamma$, and $tcg$ couplings,   as well as for their respective branching ratios, from data collected at LHC in proton-proton collisions at the energies of mass center of  7 and 8 TeV~\cite{atl12,atl16,cms16,cms17}. Specifically, data collected with the ATLAS detector~\cite{atl12}, at the energy of mass center $\sqrt{s}=7$ TeV corresponding to an integrated luminosity of 2.05 fb$^{-1}$, lead to the following branching fractions: Br$(t\to ug) < 5.7\times 10^{-5}$ and Br$(t\to cg) < 2.7\times 10^{-4}$. Later, these results where updated~\cite{atl16} at the energy of mass center of 8 TeV with an integrated luminosity of  20.3 fb$^{-1}$. A recent report on the branching fractions establishes:  Br$(t\to ug)<4.0 \times 10^{-5}$ and Br$(t\to cg)<20\times 10^{-5}$. The CMS collaboration reports upper limits on the same branching ratios, with a $95\%$ confidence level based on proton-proton collisions at a center of mass energy of 8 TeV with an integrated luminosity of 19.8 fb$^{-1}$. Thus, the bounds are  Br$(t\to c\gamma)<1.7 \times 10^{-3}$ and Br$(t\to u\gamma)<1.3 \times 10^{-4}$~\cite{cms16}. We can also mention other results reported by CMS, with $95\%$ CL, on the branching fractions for the decays $t\to ug$ and $t\to cg$,  being   $2.0\times 10^{-5}$ and $4.1\times 10^{-4}$, respectively~\cite{cms17}. This analysis is based on proton-proton collisions at $\sqrt{s}= 7, 8$ TeV, where the integrated luminosity was of 5.0 and 19.7 fb$^{-1}$, respectively.

The rest of this work has been organized as follows: In Sec. II, we describe the general renormalizable Lagrangian of neutral currents that includes both kinds of couplings, those  that violate and preserve the flavor symmetry, which is mediated by a $Z^\prime$ boson gauge. In Sec. III, we present  analytical calculations at the one-loop level for  the  $ t\to cg$ and $t\to c\gamma$ decays along with their branching ratios. In Sec. IV, the numerical results are discussed. Finally, in Sec. V, we present the final remarks.

\section{Theoretical framework}
We are interested in calculating the branching ratios for  the $t\to c\gamma$ and $t\to cg$ decays, which are induced by the $Z^\prime tc$ and $Z^\prime tu$ couplings. In this sense, we focus on the simplest extended model that predicts the existence of a new neutral massive gauge boson ($Z^\prime$): the $SU_L(2)×U_Y(1)×U^\prime(1)$ extended electroweak gauge group~\cite{plangacker,leike,Plm}. The  most general renormalizable Lagrangian that includes FCNC, mediated by the $Z^\prime$ gauge boson, can be expressed as
\begin{eqnarray}
{\cal L}_{NC}=\sum_{i,j} \left[\overline{f_{i}}\gamma^{\alpha}(\Omega_{Lf_{i}f_{j}}P_{L}+ \Omega_{Rf_{i}f_{j}}P_{R} )f_{j} + \overline{f_{j}}\gamma^{\alpha}(\Omega^{*}_{Lf_{i}f_{j}}P_{L}+ \Omega^{*}_{Rf_{i}f_{j}}P_{R} )f_{i}\right]Z^\prime_{\alpha},
\label{1a}
\end{eqnarray}

where $f_{i}$ stands for any fermion field of the SM, $Z^\prime_{\alpha}$ is the neutral gauge field associated with the gauge boson in question, and $P_{L,R}=\frac{1}{2}(1\mp\gamma_{5})$ are the chiral projectors. The $\Omega_{Lf_{i}f_{j}}$ and $\Omega_{Rf_{i}f_{j}}$ parameters represent the strength of the $Z'{f_{i}f_{j}}$ coupling. In the rest of the paper we will be assuming that $\Omega_{Lf_{i}f_{j}}=\Omega_{Lf_{j}f_{i}}$ and $\Omega_{Rf_{i}f_{j}}=\Omega_{Rf_{j}f_{i}}$.

In order to carry out calculations of the branching ratios, we consider  $Z^\prime$ gauge bosons arising from different  models, namely: the $Z_S$ of the sequential $Z$ model, the $Z_{LR}$ of the left-right symmetric model, the $Z_{\chi}$ arising from the breaking of $SO(10)\to SU(5)\times U(1)$, the $Z_{\psi}$ that emerges as a result of $E6\to SO(10)\times U(1)$, and the $Z_{\eta}$ appearing in many superstring-inspired models~\cite{AYDE}.
The Lagrangian mentioned above includes both flavor-conserving and violating  couplings mediated by the $Z^\prime$ boson. The flavor-conserving couplings, $\Omega_{Lf_{i}f_{i}}$ and $\Omega_{Rf_{i}f_{i}}$, are related with the so-called  chiral charges, $Q^{f_i}_L$ and $Q^{f_i}_L$, as $\Omega_{Lf_{i}f_{i}}= -g_2 Q^{f_i}_L$ and $\Omega_{Rf_{i}f_{i}}= -g_2 Q^{f_i}_R$ (see Table~\ref{tablaquiral}), where $g_2$ is the gauge coupling of the $Z^\prime$ boson. For the various extended models which we are interested in, the gauge couplings of the $Z^\prime$ bosons are of the form
\begin{equation}
g_2= \sqrt{\frac{5}{3}} \sin \theta_W g_1 \lambda_g, \label{1b}
\end{equation}
where $g_1=g/\cos \theta_W$ and $\lambda_g$ depends of the symmetry breaking pattern, being of the order of $\mathcal{O}(1)$~\cite{robinett,RJL}; except for the sequential $Z$ model whose gauge coupling is $g_2 =g_1$; $g$ is the weak coupling constant and $\theta_W$ is the Weinberg angle.

\begin{table}[h!]
	\center
	\caption{ Chiral-diagonal couplings of the extended models.}
	{\begin{tabular}{|c|c|c|c|c|c|c|c|}
			\hline
			Boson    & $Q^{u}_L$ & $Q^{u}_R$ & $Q^{d}_L$ & $Q^{d}_R$ &$Q^{e}_L$ & $Q^{e}_R$ &$Q^{\nu}_L$\\
			\hline
			$Z_S$    &  0.3456   & -0.1544   & -0.4228   & 0.0772    & -0.2684  &0.2316     &0.5 \\
			
			$Z_{LR}$ &-0.08493   & 0.5038    &-0.08493   &-0.6736    & 0.2548   &-0.3339    &0.2548 \\
			
			$Z_{\chi}$ &$\frac{-1}{2\sqrt{10}}$   &$\frac{1}{2\sqrt{10}}$    &$\frac{-1}{2\sqrt{10}}$   &$\frac{1}{2\sqrt{10}}$    &$\frac{3}{2\sqrt{10}}$   &$\frac{-3}{2\sqrt{10}}$    &$\frac{3}{2\sqrt{10}}$ \\
			
			$Z_{\psi}$ &$\frac{1}{\sqrt{24}}$   &$\frac{-1}{\sqrt{24}}$    &$\frac{1}{\sqrt{24}}$   &$\frac{-1}{\sqrt{24}}$    &$\frac{1}{\sqrt{24}}$   &$\frac{-1}{\sqrt{24}}$    &$\frac{1}{\sqrt{24}}$ \\
			
			$Z_{\eta}$ &$\frac{-2}{2\sqrt{15}}$   &$\frac{2}{2\sqrt{15}}$    &$\frac{-2}{2\sqrt{15}}$   &$\frac{2}{2\sqrt{15}}$    &$\frac{1}{2\sqrt{15}}$   &$\frac{-1}{2\sqrt{15}}$    &$\frac{1}{2\sqrt{15}}$\\
			\hline
		\end{tabular}\label{tablaquiral}}
\end{table}

\section{The $t\to u\gamma, c\gamma,ug, cg$ decays}
\begin{figure}[htb!]
\begin{center}
\includegraphics[width=14.0cm]{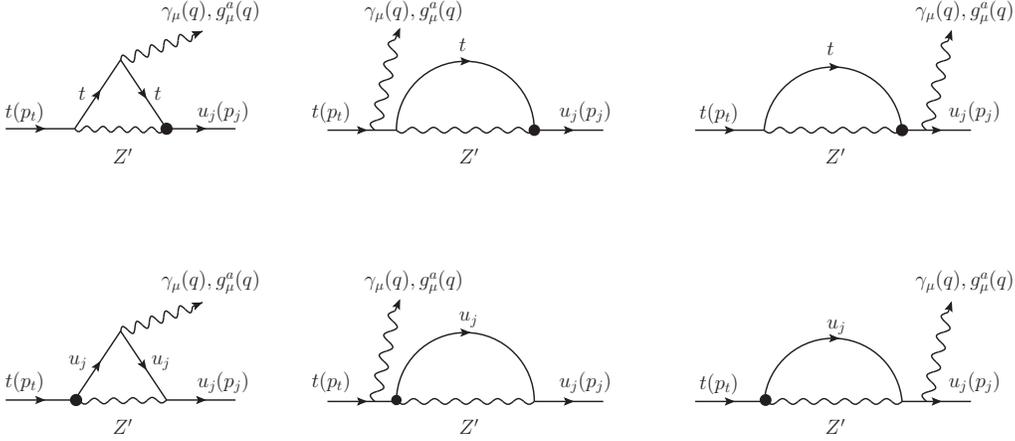}
\caption{Feynman diagrams of the $t\to u_j\gamma, u_jg$ decays.}\label{FIGDECAY}
\end{center}
\end{figure}

In  Fig.~\ref{FIGDECAY} it is shown the Feynman diagrams of the $t\to u_j\gamma, u_jg$ processes induced by a $Z^\prime$ boson at the one-loop level. The Feynman rules involved in the mentioned processes can be extracted from the  Lagrangian  in Eq. (\ref{1a}). In specific, the Feynman rules associated with the $Z^\prime t u_j$ and $Z^\prime t t$ couplings are given as $-i \gamma^{\alpha}(\Omega_{Lt u_j}\,P_L+\Omega_{R tu_j}\,P_R)$ and $i g_{2}{\color[rgb]{0,0,1}\gamma^\alpha}(Q^{t}_L\,P_L+ Q^{t}_R\,P_R)$, respectively. The calculations of the decays in question are carried out in the unitary gauge.

We take the amplitudes which comes from considering flavor violation in only one vertex, as it can be appreciated  in Fig.~\ref{FIGDECAY}.
Once the necessary calculations are done, the invariant amplitude for the $t\to u_j\gamma$ processes, being $u_j=c,u$, can be written as
\begin{eqnarray}\label{amplifoton}
\mathcal{M}(t\to u_j\gamma)=\bar{u}_j(p_j)\sigma^{\mu\nu}q_\nu\left(F_{M}^{tj}+F_{E}^{tj}\gamma^5\right)
t(p_t)~\epsilon^{*}_{\mu}(q) ,
\end{eqnarray}
where
\begin{eqnarray}\label{FFfoton1}
 F_{M}^{tj}=&\frac{ieQ_{u_k}}{64 \pi^2 m_t} \sum_{u_k=u_j,t}
\bigg[\left(\Omega_{L{tu_k}}\Omega_{Lu_ku_j}+\Omega_{Rtu_k}\Omega_{Ru_ku_j}\right)F_1\nonumber\\
&+\big(\Omega_{Rtu_k}\Omega_{Lu_ku_j} +\Omega_{Ltu_k}\Omega_{Ru_ku_j}\big)F_2\bigg],
\end{eqnarray}
\begin{eqnarray}\label{FFfoton2}
 F_{E}^{tj}=&\frac{ieQ_{u_k}}{64 \pi^2 m_t} \sum_{u_k=u_j,t}
\bigg[\left(\Omega_{Lt u_k}\Omega_{Lu_ku_j}-\Omega_{R tu_k}\Omega_{Ru_ku_j}\right)F_3\nonumber\\
&+\left(\Omega_{R tu_k}\Omega_{Lu_ku_j} -\Omega_{Ltu_k}\Omega_{Ru_ku_j}\right)F_4\bigg].
\end{eqnarray}
As usual, $e$ represents the electric charge of the electron. The different form factors $F_1$, $F_2$, $F_3$ and $F_4$ depend on the  Passarino-Veltman functions and they are explicitly  given in the appendix. Notice that the amplitude in Eq. (\ref{amplifoton}) is finite and  gauge invariant. In addition, it should be noted that the amplitude satisfies explicitly the Ward identity, since the amplitude only contains dipolar transition terms which involve the anti-symmetric tensor matrix $\sigma_{\mu\nu}$.   The $t\to u_j g$ decay has a similar structure if it is used $g$ instead of $\gamma$, therefore it would be only necessary to replace $eQ_{u_k}$ with $g_sT^a$.

The decay width for $ t\to u_jV$  is
\begin{eqnarray}\label{ancuchura1}
\Gamma(t\to u_jV)=\frac{1}{16 \pi m_t} |\mathcal{M}(t\to u_jV)|^2
\left(1-\frac{m_j^2}{m_t^2}\right),
\end{eqnarray}
where $m_t$ ($m_j$) is the mass of the  $t$ ($u_j$) quark. So, the branching ratio of the $t\to u_j V$ decay is
\begin{eqnarray}\label{6f}
\mathrm{Br}(t\to u_j V)=C_F\frac{m_{t}^3}{8\,\pi}\left(1-\frac{m_j^2}{m_t^2}\right)^3 \left(|F_M^{tj}|^2+|F_E^{tj}|^2\right)  \frac{1}{\Gamma_{t}},
\end{eqnarray}
being $C_F=1\:\:(4/3)$ for $\gamma$ $(g)$ and $\Gamma_{t}$ is the total width of the $t$ quark decay.

\section{Discussions and results}
We  numerically  evaluate the branching ratios of the $t\to u\gamma, c\gamma, ug, cg$ decays in the context of extended models by using the equation (\ref{6f}). The input values for the parameters are $\Gamma_t=1.41$ GeV, $m_t=173$ GeV, $m_c=1.275$ GeV and $\sin^2\theta_W=0.23122$ \cite{PDG}. The $Q^u_L$ and $Q^u_R$ values are given in Table I. For the numerical calculation of the Passarino-Veltman scalar functions, we use the LoopTools package~\cite{THAN}. In addition, we have also used the values $|\Omega_{tc}|=0.121596$ and $|\Omega_{uc}|=1.47856\times 10^{-3}$ both calculated at $m_{Z^\prime}=4.5$ TeV, which are  updated values with respect to the estimate obtained for the $Z^\prime tc$ and $Z^\prime uc$ couplings in a previous work~\cite{phys}.

\begin{figure}[h!]
\hspace{-2.5cm}
\begin{minipage}[c]{7cm}
\centering
\includegraphics[width=10cm]{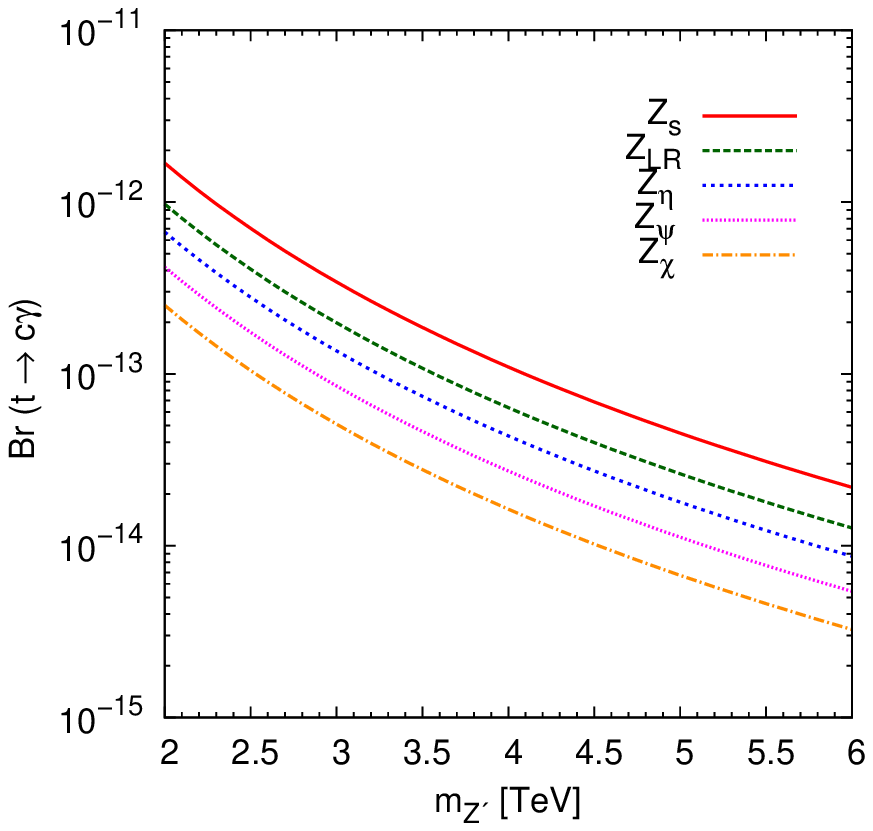}\\
{\small (a)}
\end{minipage}
\hspace{6mm}
\begin{minipage}[r]{7cm}
\centering
\includegraphics[width=10cm]{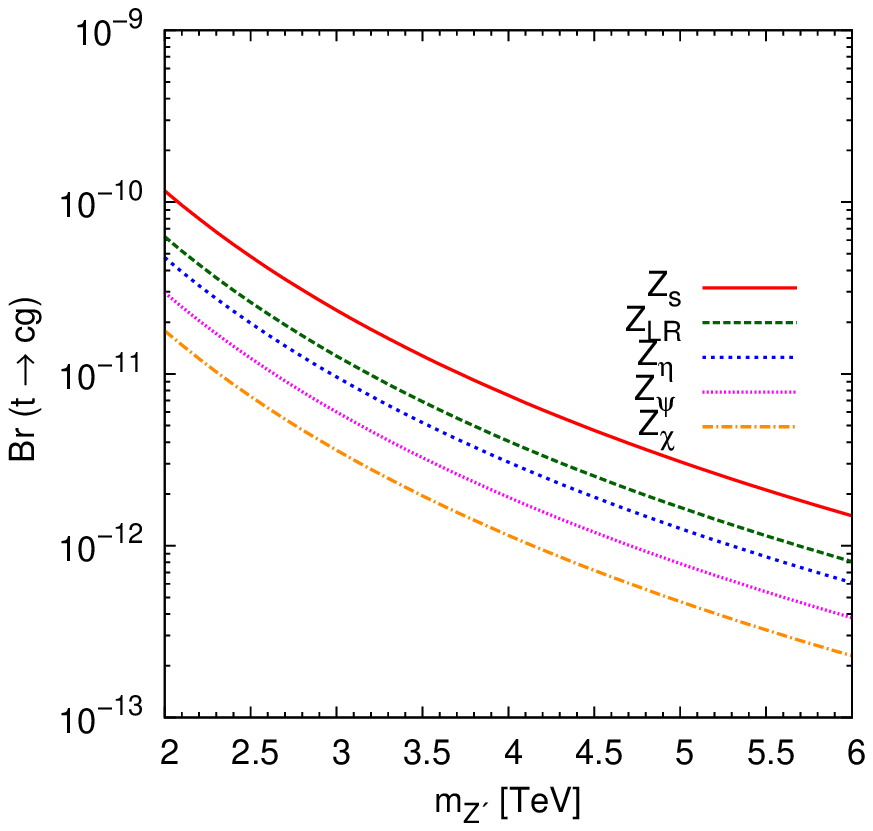}\\
{\small (b)}
\end{minipage}
\caption{Br$(t\to cV)$ as function of $m_{Z^\prime}$, where $V=\gamma,g$}\label{de3}
\end{figure}

The behavior of Br$(t\to c\gamma)$ and Br$(t\to cg)$ as a function of the $Z^\prime$ gauge  boson mass for different models is shown in Fig. \ref{de3}(a) and Fig. \ref{de3}(b), respectively. Notice that the maximum value for the Br$(t\to c\gamma)$ is of the order of $10^{-12}$, it corresponds to the $Z_S$ boson, and appears at $m_{Z^\prime}=2$ TeV. For this particle, the estimated branching ratio for the $t\to c\gamma$ process is two orders of magnitude bigger than the corresponding value found in the SM~\cite{JAS, geli}. We can also appreciate that for masses in the  interval $m_{Z^\prime}=[2,4]$ TeV, the branching ratio is one order of magnitude bigger than the corresponding value estimated in the SM and is of same order for masses $m_{Z^\prime}>$ 4.5 TeV. In relation to the other bosons: $Z_{LR}$, $Z_{\eta}$, $Z_{\psi}$ and $Z_{\chi}$, hinging on the particle, the Br$(t\to c\gamma)$ varies up to in three orders of magnitude in the analyzed mass interval. The models that offer discouraging predictions correspond to the $Z_{\psi}$ and $Z_{\chi}$ bosons. On the contrary, the $Z_{LR}$ and $Z_{\eta}$ bosons give better results, by one order of magnitude bigger than the corresponding estimates  of the SM or at least at the same order, depending of the $Z^\prime$ boson mass considered. As far as the Br$(t\to cg)$ is concern, whose behavior is shown in Fig.\ref{de3} (b), we can see that, again, the $Z_S$ boson gives the better prediction or at least comparable with respect to the SM. In fact, for $m_{Z^\prime}=$ 4.5 TeV or larger, Br$(t\to cg)\sim 10^{-12}$. The others gauge bosons that offer similar results, yet slightly lower than the $Z_S$ one correspond to the $Z_{LR}$ and $Z_{\eta}$ bosons. Once again, the $Z_{\psi}$ and $Z_{\chi}$ bosons predict a branching ratio lower than the SM on a wide range of the $Z^\prime$ mass.

\begin{figure}[h!]
\hspace{-2.5cm}
\begin{minipage}[c]{7cm}
\centering
\includegraphics[width=10cm]{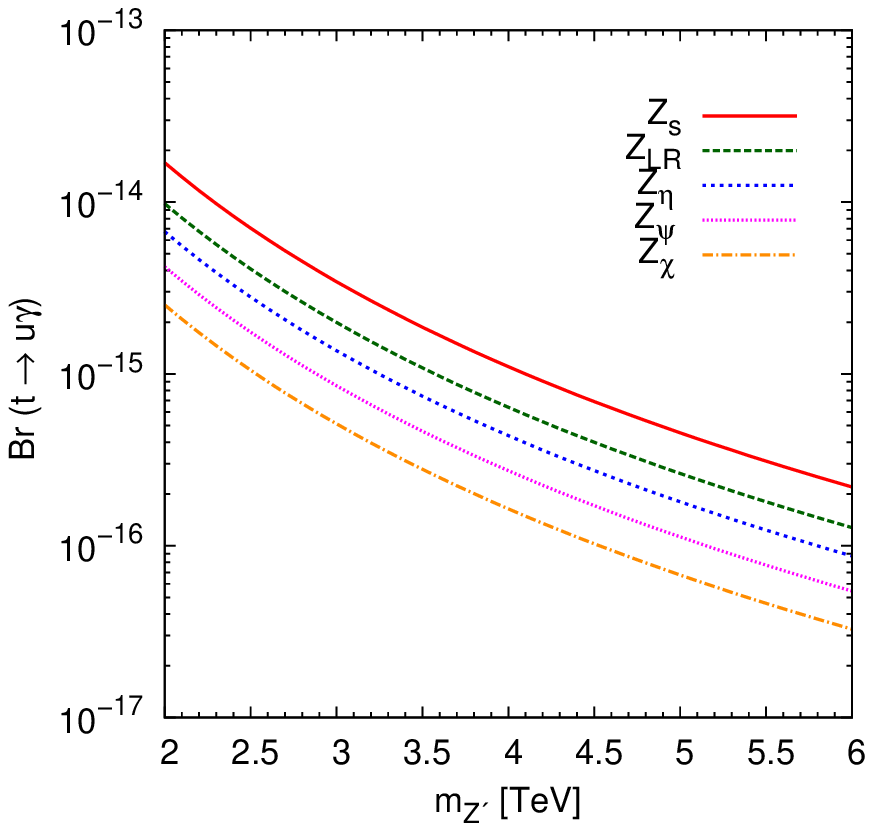}\\
{\small (a)}
\end{minipage}
\hspace{6mm}
\begin{minipage}[r]{7cm}
\centering
\includegraphics[width=10cm]{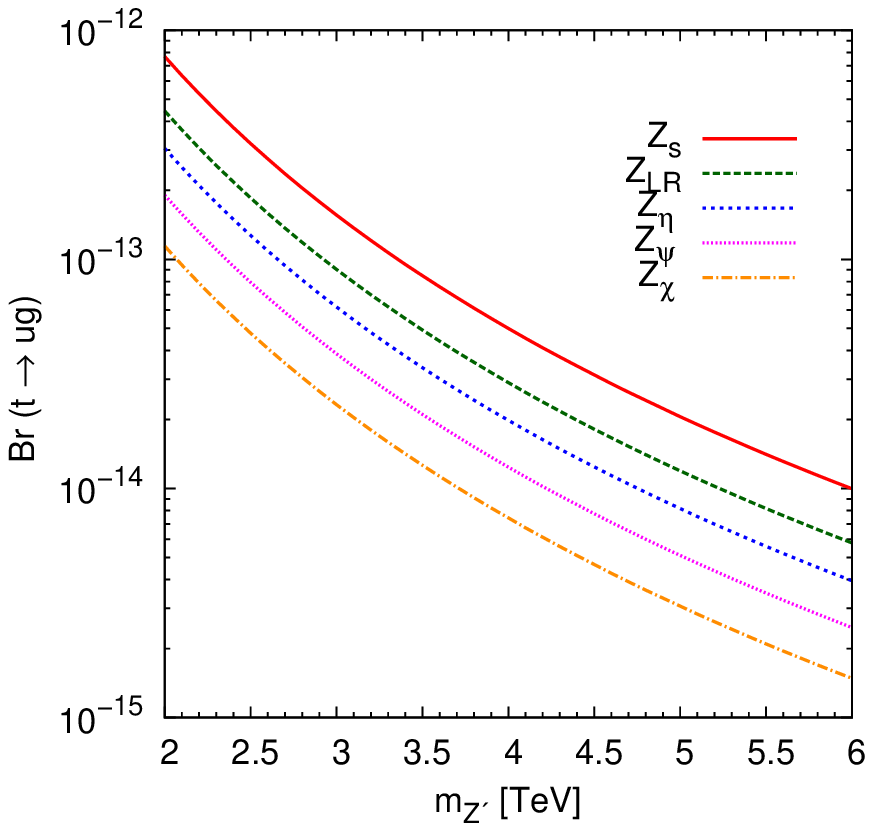}\\
{\small (b)}
\end{minipage}
\caption{Br$(t\to uV)$ as function of $m_{Z^\prime}$, where $V=\gamma,g$}\label{u1}
\end{figure}

The branching ratio corresponding to the $t\to u\gamma$ and $t\to ug$ processes can be obtained by using the previous results in Eq. (\ref{ancuchura1}), this is because the branching ratio of the $t\to c\gamma$ and $t\to cg$ decays have the same structure and it would only be necessary to change $m_j$ by $m_u$. For the analysis of the branching ratios in consideration, we use some of the data already presented in the $t\to c\gamma$ and $t\to cg$ analysis, the new inputs to be used are: $ m_u=0.0022$ GeV and the coupling parameter $|\Omega_{tu}|=1.21\times10^{-2}$ for $m_{Z^\prime}=4.5$ TeV~\cite{phys}. The  plots of the  Br$(t\to u\gamma)$ and Br$(t\to ug)$ as a function of the $Z^\prime$ boson mass  are shown in Figs. \ref{u1}(a) and \ref{u1}(b), respectively. As it can be seen, the branching ratios are two orders of magnitude more suppressed when compared with Br$(t\to c\gamma,cg)$ for all the analyzed  $Z^\prime$ bosons along the mass interval considered. As before, the model that presents better results for both decay widths is the sequential $Z_S$ model. The  branching ratios for this case, within a wide range of the mass, are bigger than the respective ones predicted by the SM and, in the worst scenario is of the same order of magnitude for  $m_{Z^\prime}=6$ TeV. In Table \ref{TABLE-bran} we show  the numerical values of the branching ratios just at  $m_{Z^\prime}=4.5$ TeV, for both decays in discussion and for all the models studied. For comparison purposes, in Table \ref{tableFrac} we present the results for the different decays in question in the context of several extended models.

\begin{table}[!ht]
	\small\center
	\caption{Branching ratio of the $t\to u\gamma,ug$ decays in extended models for $m_{Z^\prime}=4.5$ TeV.}{
		\begin{tabular}{|c|c|c|c|c|c|}
			\hline
			\multirow{2}{*}{Decay} & \multicolumn{5}{c|}{Bosons} \\
			\cline{2-6}
			& $Z_S$   & $Z_{LR}$ &$Z_{\chi}$ &$Z_{\psi}$ &$Z_{\eta}$ \\
			\hline
			$t\to u\gamma$ & 6.9004$\times 10^{-16}$ & 4.0010$\times 10^{-16}$ & 1.0266$\times 10^{-16}$ & 1.7110$\times 10^{-16}$ &2.7378$\times 10^{-16}$\\
			$t\to ug$ & 3.1293$\times 10^{-14}$ & 1.8144$\times 10^{-14}$ & 4.6558$\times 10^{-15}$ & 7.7598$\times 10^{-15}$ &1.2416$\times 10^{-14}$\\
			\hline
		\end{tabular}\label{TABLE-bran}}
\end{table}

\begin{table}[h!]
	\center
	\caption{The branching ratios of the $t\to c\gamma, cg, u\gamma, ug$ decays in some extended models.}
	{ \begin{tabular}{|c|c|c|c|c|c|c|c|c|}
			\hline
			Decay        & THDM                & MSSM      & RVSSM       & LHMT                  & LRSSM      & EDM  \\\cline{1-7}		
			$t\to c\gamma$ & $10^{-12}-10^{-6}$  &$10^{-8}$  & $10^{-5}$  & $10^{-7}$, $<10^{-9}$ & $10^{-6}$  &$10^{-6}$, $10^{-10}$ \\
			$t\to cg$      & $10^{-8}-10^{-4}$   &$10^{-6}$  & $10^{-3}$  & $10^{-2}$, $<10^{-8}$ & $10^{-4}$  &$10^{-5}$\\
			\hline
			\multirow{1}{*}{} & \multicolumn{6}{c|}{Effective Lagrangian Model} \\
			
			\hline
			\multirow{1}{*}{$t\to u\gamma$} & \multicolumn{6}{c|}{$<4\times 10^{-4}$} \\
			\multirow{1}{*}{$t\to ug$}      & \multicolumn{6}{c|}{$<2\times 10^{-3}$} \\
			\hline
		\end{tabular}\label{tableFrac}}
\end{table}

Let us mention that it could be interesting to analyze other decays similar to the ones we have studied such as the $t\to cZ, uZ, ch, uh$ decays ($h$ is the SM Higgs boson). However, in order to compute the respective amplitudes we need to use a similar procedure to the GIM mechanism, for the $\Omega$ parameters, to obtain finite amplitudes, but it results in a severe suppression to the respective branching ratios.

\section{Final Remarks}
Due to its possible connection with physics beyond the standard model, the search for flavor changing neutral currents that involves rare decays of the top quark is one of the current goals at LHC and also, in general, a topic of interest in particle physics.
In this work we have studied  the  $t\to u_j\,V$ ($u_j=c, u$, and $V=\gamma, g$) rare decays of the top quark mediated by a  neutral massive gauge boson, $Z^\prime$, predicted in several extensions of the standard model. Our theoretical framework was established on  a general yet simple renormalizable Lagrangian based on  the $SU_L(2)×U_Y(1)×U^\prime(1)$ extended electroweak gauge group, that includes flavor changing neutral currents. For our study, we considered five models, from which the $Z^\prime$ gauge boson arises, namely: the  Sequential $Z$ model, the Left-Right Symmetric model, the  model arising from the $SO(10)$ group when it is broken into the $SU(5)\times U(1)$ group, the  model that results from the breaking of $E_6\to SO(10)\times U(1)$ and the model that appears from diverse superstring-inspired models. In order to calculate the decays, we resort to the results from a previous work in which the strength of the  $Z^\prime tc$ and $Z^\prime tu$ couplings were estimated through the $D^0-\overline{D^0}$ mixing system.

Although the $t\to c \gamma$ and  $t\to c g$ decays have been studied in several models, such as the SM, the THDM, the MSSM or the so called $\mathbf{331}$ models, we found very few  reports in the literature for the $t\to u \gamma$ and  $t\to u g$ decays. In this sense, we can say that our estimation for the  branching ratios of the mentioned decays comes to extend the information about the $t\to u \gamma (g)$ decays. Regardless the degree of suppressing of the respective branching ratio, an estimation of it could  be of worth if it is competitive with respect of that found in the SM. That is the case of our estimation for a wide range of the mass of the $Z^\prime$ gauge boson, which was taken as a free parameter.

According to our calculations, the branching ratio of the decays in question grows up as the mass $m_{Z^\prime}$ becomes smaller. On the other hand, the experimental bounds for this mass is $m_{Z^\prime}>$ 3.8-4.5 TeV, corresponding to the  $Z_S$, $Z_{\chi}$ and $Z_{\psi}$ bosons; with not recent experimental reports for the $Z_{LR}$ and the $Z_{\eta}$ ones. The best results for the branching ratio of the processes studied here are provided for the $Z_S$ boson followed by the $Z_{LR}$ and the $Z_{\eta}$ bosons, respectively.

Finally, we would like to mention that the recent experimental results on the search for a new neutral gauge boson by the CMS and ATLAS collaborations indicates that the mass of a $Z^\prime$ is large than 3.9 TeV, depending on the model used. Thus, it seems that the $Z^\prime$ physics at the LHC, in the last stage of operation, has a low chance of  being observed, due to  its decoupling effects.  Moreover, in relation to the Compact Linear Collider, at last stage of its operation should be very difficult to observe $Z^\prime tc$ and $Z^\prime tu$ couplings since the collisions at the center of mass will be at 3 TeV [50], which is below of the energy for the $Z^\prime$ production.

\section*{Appendix}
The form factors $F_1$, $F_2$, $F_3$ and $F_4$
\begin{align}\label{gra1}
	F_1&=\frac{1}{(m_t-m_j)(m_t+m_j)^2 m^2_{Z^\prime}}\bigg[ m_t^2m_j(m_tm_j + 3m_k^2 - 4m_{Z^\prime}^2)(B_2 -B_3)\nonumber\\
	&+(m_k^2 - m_{Z^\prime}^2)(m_k^2+2m_{Z^\prime}^2)\bigg(-2m_t(B_2-B_3)+m_j(B_1-B_2)-\frac{m_t^2}{m_j}(B_1-B_3)\bigg)\nonumber\\
	&+m_t^3\bigg(-m_k^2(B_1-2B_2+B_3)+m^2_{Z^\prime}(B_1 - 4B_2 + 3B_3)\bigg)\nonumber\\
	&+ m_tm_j^2\bigg(m^2_k(B_1+B_2-2B_3)-m^2_{Z^\prime}(B_1 +3B_2 - 4B_3)\bigg)\nonumber\\
	&+ 2m_tm_k^2(m_t^2 - m_j^2)\bigg(m_j^2 - m_k^2- 2m_{Z^\prime}^2+ m_t(m_t + m_j)\bigg)C_0\nonumber\\
	&- m_t(m_t^2 - m_j^2)(m_k^2 + 2m_{Z^\prime}^2 + m_tm_j)\bigg],\\ \\
	F_2&=\frac{1}{(m_t^2-m_j^2) m^2_{Z^\prime}}\bigg[m_k(m_k^2-m_{Z^\prime}^2)\bigg(\frac{m^2_t}{m_j}(B_1-B_3)-m_j(B_1-B_2)\bigg)\nonumber\\
	& +m_tm_k \bigg(3(B_2-B_3)(2m_{Z^\prime}^2-m_tm_j)-(m_t^2 - m_j^2)(2m_tm_jC_0-1)\bigg)\bigg],\\ \\
	F_3&=-F_1(m_t\rightarrow -m_t), \\
	F_4&=F_2(m_t\rightarrow -m_t),
\end{align}
where
\begin{align}
	B_1&=B_{0}(0, m_k^2, m_{Z^\prime}^2),\\
	B_2&=B_{0}(m_t^2, m_k^2, m_{Z^\prime}^2),\\
	B_3&=B_{0}(m_j^2, m_k^2,m_{Z^\prime}^2),\\
	C_0&=C_{0}(m_t^2, m_j^2, 0, m_k^2, m_{Z^\prime}^2, m_k^2).
\end{align}

\section*{Acknowledgments}
J. Montaño, thanks to CONACYT for support, Cátedras, project 1753.

\end{document}